\documentstyle[sprocl,epsfig]{article}

\begin{document}
\title{PARTON CASCADES IN HIGH ENERGY NUCLEAR COLLISIONS}
\author{BERNDT M\"ULLER \\
	Department of Physics \\
	Duke University \\
	Durham, NC 27708-0305, USA}

\maketitle 

\abstracts{
This is a review of the parton cascade model (PCM) which provides a QCD-based description of nucleus-nucleus reactions at very high energy.  
The PCM describes the collision dynamics within the early and dense phase of the reaction in terms of the relativistic, probabilistic transport of perturbative excitations (partons) of the QCD vacuum, combined with the renormalization group flow of the parton virtuality.  The current state of numerical implementations of the model, as well as its predictions for nuclear collisions at RHIC and LHC are discussed. }

\section{The Parton Cascade Model}

The parton cascade model \cite{GM92}
 (PCM) was proposed in 1991 and developed further during the following years, especially, by Geiger \cite{Gei92a,Gei92b,Gei94,Gei95}.
Its aim is to describe the energy deposition, thermalization, and chemical equilibration of matter in ultra-high energy nuclear collisions, providing a full space-time picture of the event up to the moment when individual hadrons are formed.  The model was, at least originally, not conceived as an ``event generator'' that would predict a full set of hadron momentum distributions in the final state.  The parton cascade model code VNI, developed by Geiger, allows us to make such predictions because it contains the implementation of a hadronization scheme similar to those used in jet fragmentation algorithms.  However, it should be clear that this runs somewhat counter to the original purpose of the PCM, namely, to explore the range of validity of perturbative QCD in nuclear reactions.  Nevertheless, the predictions of the PCM with an added hadronization stage are experimentally useful. 

The conceptual basis of the PCM is the inside-outside cascade model 
\cite{AKM80} of hadronic reactions, which implements the concept that new matter produced in hadronic interactions at high energy is formed outside the intersecting world-tubes of the colliding hadrons.  Bjorken's hydrodynamical model \cite{Bj83} was developed to describe the evolution of this newly formed matter in the central space-time region after thermal equilibration.  Beginning in the mid-1980's it was realized that the deposition of energy into this region may be described in terms of ideas derived from perturbative QCD (minijets) in the case of colliding heavy nuclei \cite{HK86,BM87,KLL87,EKL}.  A computer code (HIJING) incorporating some of these ideas was developed by Gyulassy and Wang \cite{GW,Wang97}.

The parton cascade model combines these concepts into one unified scheme for the description of the space-time evolution of matter in nuclear reactions.  Its three main ingredients are: 

\begin{enumerate} 
\item  
The {\em initial state} is viewed as incoherent ensemble of partons determined by the nuclear parton distribution 
 functions $q_f(x, Q^2)$ and $g(x, Q^2)$, where the subscript {\it f} denotes t
he quark flavor and $g$ stands for the gluon distribution.  $x = p_z/P$ is the longitudinal momentum fraction of the nucleon carried by the parton, and $Q^2$ is the parton ``scale'' or virtuality.  Before any interaction occurs, $Q^2$ is generally taken as space-like.  Our knowledge about the space-time structure of the nuclei before the collision and our limited information about the intrinsic transverse momenta of partons is then used to construct a model for the six--dimension phase space distributions of partons before the interaction: 
$q_f (r, p), g(r, p)$.  The parton distributions are conveniently initialized at the  scale $Q^2_0 = \langle p^2_T \rangle_{\rm coll}$ of the average momentum scale of the primary parton-parton interactions. 

\item  
The {\em time evolution} of the parton phase distributions is governed by a relativistic Boltzmann equation with ``leading-log'' improved lowest-order collision terms.  Only binary interactions are allowed, but the final state can have (and generally has) more than two particles.  As is well known, the higher-order improvement of the cross sections by means of the leading logarithmic approximation is equivalent to the scale evolution of the parton distributions according to the DGLAP equation.  Motivated by quantum mechanical considerations, the space-time picture of parton propagation before and after interactions is closely related to their off-shell propagation: the formation of a parton with virtuality scale $Q$ takes a time $\tau_f (Q) \approx \hbar/Q$. 

\item 
When the parton distributions become sufficiently dilute, they {\em hadronize}.  In VNI the hadronization is described by a clustering algorithm, followed by the decay of excited hadrons.  The transition is assumed to occur when the average virtuality of the partons falls below a critical value $Q_{\rm crit}\approx 1{\rm GeV}$, because partons no longer scatter with sufficient energy.
\end{enumerate}

From a gradient expansion of the evolution equation for the parton Wigner distribution two equations can be derived \cite{Gei96}.  The first equation 
\begin{equation} 
  p^\mu \frac{\partial}{\partial r^\mu} F_i(r, p)
= C_i (r, p)                                        \label{star} 
\end{equation}
describes the free propagation of partons which is intermittently modified by collisions given by the binary collision terms $C$.  The second equation 
\begin{equation}
  p^2 \frac{\partial}{\partial p^2} F_i (r, p) 
= S_i (r, p)                                     \label{twinkle} 
 \end{equation} 
describes the evolution of the parton distributions with respect to virtuality or ``off-shellness'' $p^2$.  This equation is a generalization of the usual mass-shell condition $F(r, p) \sim \delta (p^2-m^2)$ to the case where the on--shell particle distribution cannot be defined.  $S_i (r, p)$ describes the splitting of single off-shell partons into two partons of smaller virtuality. 

The two equations can be viewed as quantitative representation of Feynman diagrams of the type shown in Figure 1.  The collision term $C_i$ is represented by the binary collision diagram contained in the box at the heart of the complex Feynman diagram, whereas the splitting term $S_i$ is represented by the branchings diagrams of the initial-- and final-state partons.\footnote{Strictly speaking, $S_i$ describes the differential branching probability for an infinitesimal change in virtuality; the diagram is an integral representation of $S_i$.}  The Feynman diagram of Fig. 1 is finite only if the virtualities of all final--state partons are limited by some infrared cut-off $\mu^2$.  In an isolated event $\mu^2$ describes the {\em hadronization scale}, i.e. the virtuality scale below which partons can no longer be considered as approximately free, perturbation quanta.  In a dense medium, where partons rescatter often, $\mu^2$ is determined by the frequency of rescatterings (see Section 3.2). 

\begin{figure}[htb]
\vfill
\centerline{
\begin{minipage}[t]{.47\linewidth}\centering
\mbox{\epsfig{file=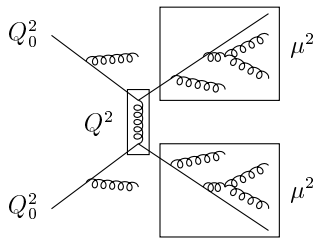,width=.9\linewidth}}
\end{minipage}
\hspace{.04\linewidth}
\begin{minipage}[b]{.47\linewidth}\centering
\caption{Graphical representation of the QCD transport equations (1,2) defining the parton cascade model.  The $2 \to 2$ scattering process at momentum scale $Q^2$ is followed by the virtuality evolution from $Q^2$ to $\mu^2$.}
\label{fig1}
\end{minipage}}
\end{figure}

In the leading-logarithmic approximation (LLA), the differential cross section described by the Feynman diagram of Fig. 1 factorizes into a product of terms for each of the two incoming and outgoing branch processes and one for the binary scattering process.  Each branching term, in turn, is represented by a product of factors describing the individual branching events and the probabilities for the partons {\em not} to branch further in between.  In other words, the Feynman diagram of the type shown in Fig. 1 defines a Markov process, as it would be expected for any process that can be described by a probabilistic one--body transport equation.  This statement is no longer valid, if one tries to go beyond the LLA.  However, certain effects beyond the LLA can still be described in terms of conditional probabilities, such as angular or $k_T$--ordering of limited gluons and soft--gluon interference effects.  These effects are quantitatively important and have been incorporated in parton cascade codes \cite{Gei92b}.  

Xiong and Shuryak \cite{XS94} have argued that one should include all $2 \to n$ parton tree diagrams in the collision term of (\ref{star}), instead of relying on the LLA.  Of course, if one does this, it is necessary to set the right--hand side of (\ref{twinkle}) equal to zero, lest one double--counts parton splittings.  However, it is important to note that parton splittings are related by unitarity to loop diagrams that describe the running of the strong coupling constant $\alpha_s (Q^2)$: 

Both splittings and $\alpha_s$--running are described consistently in the LLA, which therefore satisfies the unitarity condition.  In plain terms, the combined probability for all $2\to n$ parton diagrams with $n \geq 3$ reduces the probability for the occurrence of a $2 \to 2$ scattering and so on. 
By summing all $2 \to n$ free diagrams, but not including the associated loop diagrams, unitarity is violated.  
This leads to the divergence of the sum over $n$ for rather small c.m. energies, even in the presence of an infrared cut--off for the internal propagators.  If one naively includes the full sum at energies below this artificial divergence, one finds \cite{XS94} that the chemical equilibration of parton distributions is driven by processes with many final--state gluons, i.e. by $2 \to n$ scatterings with $n \geq 3$.  In my opinion, this result indicates that this approach is questionable.  If one includes scattering processes with more than two quanta in the final state, one also needs to account for unitarity corrections due to loop diagrams.  This is especially important in the presence of a medium, where loop diagrams are enhanced by in--medium contributions (see Section 3.1). 

In the following I will discuss two important issues:  
\begin{enumerate} 
\item 
The space--time picture governing the initial--state parton distributions.  The issue is closely connected with the problem of the decoherence of the initial parton wavefunctions. 

\item 
The problems of infrared divergences of the perturbative parton cross sections.  This issue is closely related to in--medium corrections to these cross sections, as well as to coherence properties of the initial--state wavefunctions.
\end{enumerate} 
The approach to local thermal equilibrium has been extensively studied within the framework of the parton cascade picture.  \cite{Wang97,EW94,HW96} Without repeating the detailed arguments here, the PCM approach predicts a very short kinetic equilibration time, $\tau_{\rm th} \ll 1$ fm/c, which is confirmed by full numerical calculations.\cite{Gei92a}

\section{Initial--state space--time picture} 

The probabilistic interpretation of parton distributions measured in deep--inelastic scattering is based on a summation over all final hadronic states.  A similar interpretation of one--body distributions arising in transport theory is based on the low--level truncation of the (BBGKY) hierarchy of Green functions and on an expansion in power of $\hbar$.  The validity of this picture ultimately relies on the separation of time scales in a dynamic process.  Although these issues are generally well known  \cite{MR96}, their full implications for nuclear parton cascades have not been explored.  Recent advances \cite{Baier} in our understanding of multiple scattering in QCD have shed some light on the intricacies of the formation time concept in non--abelian gauge theories, but more needs to be understood.

The original parton cascade model relied on some basic assumptions about initial parton distributions in space--time. \cite{EW94}  Denoting the parton light--cone momentum by $p^+$, the parton distributions were assumed to be distributed longitudinally according to the uncertainty relation: $\Delta p^+ \Delta x^- \geq \hbar$.  Soft partons have the widest distributions in the variable $x^-$, and were assumed to travel both ahead and behind the Lorentz contracted valence quark distributions.  The argument was that this would not violate causality, because soft partons are emitted at a long distance before the collision and, travelling at the speed of light, can arrive significantly ahead of the quarks that emitted them. 

The space--time picture of soft partons has been put on a much firmer foundation in recent years by the work of McLerran, Venugopalan and others on the random light--cone source model \cite{MV94} (RLSM).  In this model, one views the valence quarks constituting the fast-moving nucleus as a thin, Lorentz contracted sheet of locally random color sources.  The color source is locally random, because valence quarks from several nucleons contribute at the same point in transverse space.  The area density of color sources is given by 
$\mu^2 = 3A/\pi R^2$, where $A$ is the nucleon number and $R$ is the nuclear radius.  Clearly $\mu \approx A^{1/6}$, hence $\mu$ becomes a large scale\footnote{In practice, $\mu \leq 1$ GeV even for the heaviest nuclei.}
 for sufficiently heavy nuclei, 
and $\alpha_s (\mu^2) \ll 1$ can serve as basis for a new type of perturbative expansion.  Formally, the model maps into the problem of weakly coupled QCD in the presence of a random two--dimensional color source. 

As shown by Kovchegov \cite{Kov96}, this model can be rigorously derived by standard light--cone techniques, which permit an explicit representation of the Gaussian ensemble of color sources.  This representation can also be used to calculate the perturbative emission of soft gluons in collisions between two nuclei, described as collision between two sheet-like clouds of valence quarks.  
\cite{KR96,RMM97}  At leading order this soft gluon radiation is given by:
$$
\frac{dN_g}{dy d^2k_ \perp d^2b} 
 = \frac{ 4\alpha^3_s}{\pi^2 k^2_\perp} 
\frac{N^2_c -1}{N_c} \langle T_{AB}(b) \rangle 
 \int d^2g  
\frac{ F(qa) F(|k-g| a)}{ q^2(k-q)^2}
$$
where $F(qa)$ is the color dipole form factor of the nucleon and $T_{AB}(b)$ denotes the nuclear profile function.  It can be shown \cite{GML97} that this classical gluon radiation smoothly matches onto the perturbative minijet production of gluons at higher $k_\perp$. 

Going beyond the classical approximation by including gluon loop diagrams leads to a better and more rigorous understanding of the space--time distribution of soft gluons in a heavy nucleus.  \cite{KMW97}  The quantum corrections can be formulated in the framework of a space--time analogue of the renormalization group equations, describing the cascade of gluon emission leading to a power--law enhancement of soft gluons similar to the BFKL equation. \cite{KLW97} 
The RLSM approach also describes saturation effects in the parton distribution at small $k_\perp$.  It would be very interesting to study its relation to the perturbative theory of gluon shadowing \cite{Qiu,EQW} in this regime.  

The picture that emerges is the following: Gluons in the classical field generated by the valence quarks are fully Lorentz contracted by the Lorentz factor $\gamma$ associated  with the colliding nuclei, but gluons spawned by a splitting of those primary gluons experience only a partial Lorentz contraction of order 
$x\gamma$, where $x$ is the momentum fraction carried by the parent gluon.  As the branching process evolves to softer and softer gluons, the spatial extent of this gluon cloud becomes more and more diffuse in the light--cone variable $x^-$.  This result confirms the intuitive picture embodied in the original parton cascade model, and provides a quantitative formulation of it. 

\section{Medium effects} 

Medium--effects on parton--parton interactions are essential for the viability of the parton cascade model.  To wit, the application of the PCM to nucleon--nucleon collisions requires the introduction of {\em ad hoc} cut--offs describing {\em nonperturbative} QCD effects, such as quark confinement and chiral symmetry breaking.  Medium effects, which grow rapidly in size as function of $A$, can produce {\em perturbative} cut--offs when the density of the medium becomes sufficiently high.  E.g., QCD is known to become perturbative \footnote{Some nonperturbative effects remain even at high $T$, precisely because static magnetic interactions are not screened by perturbative in--medium interactions.} at high temperature when the electric screening mass $\mu_{\rm D} \gg \Lambda _{\rm QCD}$.  

Thus, medium effects work in favor of the parton cascade model.  The problem is that medium effects are very complicated and not easily treated correctly.  The two main medium effects that are known to provide effective infrared cut--offs to perturbatively divergent parton interactions are: 

\begin{itemize}
\item 
(color--electric) screening, which suppresses soft $2\to 2$ scattering amplitudes; 

\item 
(gluon) radiation suppression, which dampens $2 \to 3$ (and $2 \to n)$ amplitudes with soft particles in the final state.  
\end{itemize} 
 Dynamical screening, at lowest order, is described by the in--medium contributions to the one-loop gluon polarization function: 
$$
\centerline{\mbox{\epsfig{file=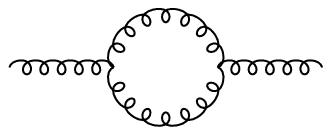}}  }
$$
At moderately high $q_\perp$, the gluon population grows like $n(k) \approx (A_1 A_2)^{1/3}$, providing a screening scale $\mu(A)$ that increases rapidly with nuclear size.   \cite{BMW92}

Radiation suppression, also known as the Landau-Pomeranchuk-Migdal (LPM) effect, is a much more complicated effect.  Its theoretical description requires a good understanding of the multiple scattering problem in QCD.  Considerable progress has recently been made in this area, especially through the work of Baier et al. \cite{Baier}, Zakharov 
\cite{Zak} and others \cite{KHN}.  The main difference between QCD and the well-known case of QED is that a radiated gluon also rescatters in the medium at the same order in $\alpha_s$ as the emitting particle; this is not so for a photon in an electromagnetic plasma. 

Revisiting the diagram shown in Fig.\ 1, the parton cascade model requires infrared cut-offs for both the central $2 \to 2$ scattering matrix element and each of the four branching cascades.  This is where medium effects help: at high density one expects the in--medium modifications of the elementary scattering amplitude to ensure an infrared safe behavior.  To date, two attempts have been made to practically implement the action of these medium effects: 
\begin{enumerate} 
\item[(1)]
In the self-screened parton cascade model \cite{EMW96} (SSPCM), the color-electric screening scale $\mu(p_T)$ was calculated self--consistently for primary parton interactions only, and the further evolution of the parton plasma was described in the framework of the hydrodynamical model.

\item[(2)] 
In the PCM code VNI \cite {VNI}, the space-time picture of parton interactions is linked with the virtuality evolution of partons.  A new interaction is permitted if its momentum transfer exceeds the virtuality scale of the participating partons at that time.  Soft--radiation is suppressed in the medium by this rule, because intermediate collisions continuously reset the parton virtuality. 
\end{enumerate} 
Both of these approaches are based on specific assumptions about the relation between the space-time and virtuality evolution of off-shell components of the parton distributions.  One way to study this issue vigorously is the Wigner function representation.  First results \cite{BD98} obtained by this method are interesting, but do not yet fully address the complications encountered in a QCD parton cascade.  A different approach to the problem of the space-time picture of off-shell quantum fluctuation is based on a modification of the QCD evolution equations \cite{MS98} to include an infrared scale.  In free space this infrared scale is determined by properties of the final state (hadronization scale); in a medium it is determined by screening effects. 

\subsection{Self--screened parton cascade} 

Here one considers the scattering of an initial state parton as completed after a time $\tau(p_T)$ which depends on the momentum transfer in the reaction.  The uncertainty relation suggests $\tau(p_T) \sim \hbar /p_T$.  (We will drop the factor $\hbar$ in the following.)  The scattered partons are then assumed to screen the scattering processes that involve a smaller momentum transfer:
$$
\mu^2_{\rm D}(p_T) 
= \frac{3}{\pi^2} \alpha_s (p^2_T) \int^\infty_{p_T}  d^3k 
|\nabla_k n(k)|.
$$
The density of partons scattered at $p_T$ is, in turn, influenced by the screening because the differential cross section depends on $\mu_0$: 
$$
\frac{d\hat{\sigma}}
      { dp^2_T}     \sim 
\frac{\alpha_s (p_T)^2} 
     {(p^2_T + \mu^2_{\rm D}(p_T))^2}  \left|M(\hat{s}, \hat{t})\right|^2.
$$
If $\mu_{\rm D}(p_T)$ becomes large enough at low $p_T$, so that $d\hat{\sigma}/dp^2_T$ remains perturbatively small, the coupled set of equations can be integrated down to $p_T = 0$.  Since the rapidity density of scattered partons grows as $(A_1A_2)^{1/3} \ln \sqrt{s}$, this condition requires large $A$ and high energy.  The SSPCM concept is closely related to the RLSM approach proposed by McLerran and Venugopalan \cite{MV94}. 

Quantitatively, one finds that $\mu_{\rm D}$ approaches about 1 GeV at low $p_T$ in Au + Au collisions at RHIC energy (100 GeV/u) and 1.5 GeV at LHC energy (2.75 TeV/u).  The differential minijet cross section as function of $p_T$ peaks at about the same value, clearly showing the improved infrared behavior of the self--screened parton cascade.  The total deposited energy within one unit of rapidity and after a characteristic formation time of 0.25 fm/c is $\epsilon_0 \approx$ 60 GeV/fm$^3$ (RHIC) and $\epsilon_0 \approx $ 430 GeV/fm$^3$ (LHC).
  The conditions established by the SSPCM can, therefore, be taken as initial conditions for the thermal and chemical evolution of a quasi--equilibrated parton plasma. 

The equations for the evolution of such a plasma were formulated by Bir\'o et al. \cite{Biro93} and by Xiong and Shuryak \cite{XS94}.  Extensive calculations \cite{Stri94,SMM97}, including longitudinal and transverse expansion, have shown that the plasma cools down to the critical temperature of QCD ($T_c \approx 150$ MeV) after 5 fm/c (RHIC) and 10 fm/c (LHC).  The emission of electromagnetic probes by such an evolving QCD plasma has also been calculated.  \cite{SMM97}

\subsection{Monte-Carlo space--time cascade} 

The statistical implementation of a parton cascade by means of a Monte-Carlo code (VNI \cite{VNI}) achieves an improved infrared behavior through heuristic rules that suppress certain interactions on the basis of kinematic considerations.  The first rule asserts that independent scattering events involving the same parton require a sufficient time separation so that the time between scatterings is larger than the duration of the individual events.  With the duration of an interaction again defined as $\tau (p_T) \sim p^{-1}_T$, where $p_T$ is the momentum exchange, this requires that the time between interactions $\Delta \tau > \tau (p_T)$.  Another way of ensuring this condition is to endow a parton after a scattering by $p_T$ with an initial virtuality $Q_0 = p_T$, which then gradually decreases with time as $Q(\tau) = Q_0 \tau(p_T)/\tau$.  A subsequent scattering with $p^\prime_T$ requires that $p^\prime_T > Q(\tau)$ at the moment of the interaction.  A second similar rule suppresses soft parton splittings in the presence of multiple scattering.  Again, this rule can be formulated in terms of a parton virtuality that decreases with time between scatterings and is reset by each new interaction (see Fig.\ 2).

\begin{figure}[htb]
\vfill
\centerline{
\begin{minipage}[t]{.47\linewidth}\centering
\mbox{\epsfig{file=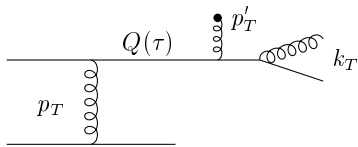,width=.9\linewidth}}  
\end{minipage}
\hspace{.06\linewidth}
\begin{minipage}[b]{.47\linewidth}\centering
\caption{Illustration of in-medium suppression effects incorporated in the VNI code.  The virtuality of a scattered parton evolves with time, $Q(\tau)$.  Sequential scatterings or branchings are suppressed, if $Q(\tau)$ is too large.}
\label{fig3}
\end{minipage}}
\end{figure}

Although VNI still contains ``arbitrary'' infrared cut--off parameters (determined by a comparison with nucleon-nucleon interactions), these are needed to limit soft scatterings in the initial set of parton interactions, and they become important toward the end of the cascade evolution when the parton plasma becomes more and more dilute.  The cut--off parameters effectively determine the end of the cascade evolution and, by suppressing soft interactions, they trigger the hadronization of parton clusters. 

The fact that this improved parton cascade is much less dependent on arbitrary cut--off parameters makes it possible to explore applications of this model at lower energies where nucleon-nucleon collisions provide little information.  Recently, Geiger and Srivastava \cite{GS97} have studied the predictions that the VNI code makes for nuclear collisions in the energy regime of the CERN-SPS.  While most of the particle yield at rapidities $|y| \geq 2$ is produced by fragmentation of the unscattered beam remnants, the model predicts a significant contribution to particle production at central rapidity from partons that have undergone perturbative scattering. \cite{GM97}  This contribution is rising rapidly with nuclear mass $A$, roughly as $(A_1A_2)^{1/3}$.  This perturbative contribution to the energy deposition at $|y| \leq 1$ coincides with a rapid increase of the energy density in scattered partons at $\tau <$ 1 fm/c, which rises from about 2 GeV/fm$^3$ in S+S to 5 GeV/fm$^3$ in Pb + Pb.  This rise may be correlated with the much enhanced suppression of charmonium production in Pb + Pb collisions as observed by the NA50 experiment. \cite{NA50}

\section{Summary}

The parton cascade model was developed to provide a QCD-based description of the approach to a locally thermalized state in collisions of heavy ions in the RHIC energy regime and beyond.  In its original formulation the PCM predictions were critically dependent on several cut--off parameters that had to be determined from $pp$ collision data.  Recent advances in incorporating medium effects into the parton interactions have reduced this dependence significantly, possibly allowing the application of the PCM over a wider energy range.  Results obtained for nuclear collisions at CERN-SPS energies are intriguing. 

The in--medium effects that modify parton-parton interactions not only reduce the parameter dependence of the model, they also provide valuable insight into the dynamics of a dense parton plasma.  It is clear that we are here just at the beginning. The transport properties of off-shell quanta need to be understood much better, not only in cases where the off-shell propagator is dominated by a well-defined resonance, but especially in the case where the particles never get close to their mass shell as it applies to QCD.  Another open question concerns the need for mean color fields.  Such fields are not included in present versions of the PCM, but the random light--cone source model suggest that mean fields may be essential ingredients of a complete description of soft processes in nuclear collisions.  One may have to take an average over a Gaussian ensemble of mean fields, thus the width of the field distribution may be more important than the expectation value which remains zero.  It would be interesting to explore possible connections of the RLSM to the traditional chromo--hydrodynamical model. \cite{BC85,GKM85} 

Ultimately, the question is whether the parton cascade model can be replaced by a controlled approximation scheme where, the principle, successive orders of ever more sophisticated corrections can be calculated.  We are still far away from a consistent formulation of transport phenomena off equilibrium in QCD, and it is even unclear whether we have a full insight into what are the small parameters in such an approximation scheme.  It is clear that a high density of excitations of the QCD vacuum is an essential condition, but there are many subtleties if one wants to go beyond this statement.  However, progress in this field is steadily made, and there is reason to hope that a consistent formulation of transport phenomena in QCD can ultimately be achieved. 

My final remarks concern the treatment of the late phase of a heavy ion collision when the dense matter breaks up into individual hadrons.  Here one has the choice of either applying a phenomenological hadronization model to make the transition to a hadronic cascade, or to change from the parton cascade model to a hydrodynamical description already in the dense plasma phase as soon as approximate local (kinetic) equilibrium has been reached.  The second scheme has the advantage that an equation of state describing the QCD phase transition can be easily incorporated; it has the disadvantage that another transition to some cascade--like scheme is needed in the very late phase when the mean free paths of particles become too long to sustain the hydrodynamical description.

\section*{Acknowledgments}
This work was supported in part by grant no. DE-FG02-96ER40945 from the U.S.\ Department of Energy.

\end{document}